\documentstyle[11pt]{article}
  \newcommand {\nc}{\newcommand}
  \nc{\eq}{\begin{equation}}
  \nc{\en}{\end{equation}}
  \nc{\eqa}{\begin{eqnarray}}
  \nc{\ena}{\end{eqnarray}}
  \newtheorem{definition}{Definition}
  \newtheorem{lemma}{Lemma}
  \newtheorem{proposition}{Proposition}

  \def\etal{{\it et al} }
  \def\DCv{Dirac-Connes }
  \def\dop{Dirac operator}
  \def\dopv{Dirac operator }
  \def\rep{representation}
  \def\repv{representation }
  \def\dlt{\delta}
  \def\pr{\prime}
  
  \def\intg{{\cal Z}}
  \def\complex{{\cal C}}
  \def\Alg{{\cal A}}
  \def\Hs{{\cal H}}
  \def\Drc{{\cal D}}
  \def\unit{{\bf 1}}
  \def\prf{{\bf{Proof:}}\\}
  \def\endprf{${\bf{\Box}}$\\}
  \nc {\norm}[1]{\parallel{#1}\parallel}
  
 \title{Noncommutative Geometry of Lattice and Staggered Fermions}
 \author{Jian Dai\thanks{E-mail: jdai@mail.phy.pku.edu; Postal address: Room 2082, Building 48, Peking University, Beijing, P. R. China, 100871},
  Xing-Chang Song\thanks{E-mail: songxc@ibm320h.phy.pku.edu.cn; Postal address: Theory Group, Department of
  Physics, Peking University, Beijing, P. R. China, 100871}\\
  Theory Group, Department of Physics, Peking University
  }
 \date{December 18th, 2000\\
 Revised: April 19, 2001}
 
\begin{document}
  \noindent
  \begin{titlepage}
  \maketitle
  \begin{abstract}
   Differential structure of a d-dimensional lattice, which is essentially a noncommutative exterior
   algebra, is defined using reductions in first order and second order of universal differential
   calculus in the context of noncommutative geometry(NCG) developed by Dimakis {\it et al}. This differential structure
   can be realized adopting a \DCv operator proposed by us recently within Connes' NCG.
   With matrix \rep s being specified, our \DCv operator corresponds to staggered \dop, in the
   case that dimension of the lattice equals to 1, 2 and 4.\\

  \end{abstract}
  \end{titlepage}
  \section{Introduction}
   In our recent work \cite{ds111}, we formulated a Dirac operator on discrete abelian
   group to bridge the noncommutative geometry devised by Dimakis \etal based on {\it reduced differential
   calculus} \cite{dimakis} and spectral noncommutative
   geometry (sNCG) \cite{NCG}\cite{NCG1}, and we referred this operator as {\it \DCv operator}.
   It has no longer been a new intuition to consider lattice \dopv within the framework of noncommutative geometry (NCG).
   Feng \etal employed the intuition of ``half-spacing'' lattice in \cite{fls},
   while Vaz generalized Clifford algebra to be non-diagonal in spacetime \cite{v}; Balachandran
   \etal studied another type of Dirac operator in the context of discrete field theories upon fuzzy sphere \cite{madore} and its
   Cartesian products \cite{fuzzy}. It is worthy to be remarked that all these ideas that we mention above were more or less with the aim to
   resolve the species doubling puzzle of massless fermion on
   lattices \cite{double} which has been explored for more than two decades by
   lattice field theorists.\\

   In this contribution, we will show that under a specific matrix representation, our \DCv operator which is rooted in
   pure geometry possesses an interpretation of staggered Dirac operator emerged from lattice field theory (LFT) \cite{k-s}.
   Below we give an outline of this article. A canonical differential structure can be implemented onto a lattice as a quotient
   algebra of the universal differential calculus on this lattice by a collection of first order and second order {\it reductions}.
   This differential structure is essentially a noncommutative exterior
   algebra. Our \DCv operator provides a natural representation for this
   reduced calculus on a spinor Hilbert space and this representation naturally has the same dimension
   as that of staggered fermions (Section \ref{a}).
   After matrix \repv being assigned, the correspondence between our \DCv operator and staggered \dopv
   can be computed explicitly in cases that the dimension of
   the underlying lattice equals to one, two and four (Section \ref{b}).
   We will also discuss the relation of our formalism and that of Takami \etal \cite{japan} (Section \ref{c}).
  \section{Noncommutative Geometry on Lattice: Two Approaches}\label{a}
   A noncommutative space can be described in either quantum algebraic
   way or operator algebraic way. NCG of a discrete point set, with or without a group structure being endowed,
   has been formulated well along the first approach
   by Dimakis {\it et al} \cite{dimakis}.
   A d-dimensional lattice, being a specific object in this category, can be parametrized
   by a direct-product group $\intg^d$ where $\intg$ is the integer addition
   group, namely each element in $\intg^d$ can be labeled by one d-tuple vector $x$ whose
   components $x^i, i=1,2,...,d$ are integers. Let $\Alg$ be the algebra
   of complex functions on $\intg^d$. The group translations on $\intg^d$ being pulled back onto $\Alg$ are defined by
   $(T_xf)(y)=f(x+y), \forall x,y\in \intg^d, \forall f\in \Alg$.
   A natural linear basis of $\Alg$ is a complete class of delta-functions on this lattice $\{ e^x, x\in \intg^d: e^x(y)=\Pi_i\dlt^{x^iy^i}, \forall
   y\in \intg^d \}$.
   One can easily check that $T_xe^y=e^{x-y}$. We will use $a=1$
   as the convention for lattice constant all through this work.
   \begin{definition}\label{def1}
   Universal Differential Calculus $(\Omega_u(\Alg),d)$ over $\intg^d$:\\
   i) $\Omega_u(\Alg)=\oplus^\infty_{k=0}{\Omega_u^k(\Alg)}$ is a
   bimodule over $\Alg$ with $\Omega_u^0(\Alg)=\Alg$ and the elements in $\Omega_u^p(\Alg)$ are
   referred as p-(order)forms;\\
   ii) $\Omega_u(\Alg)$ is a $\intg$-graded algebra:
   $\Omega_u^p(\Alg)\cdot\Omega_u^q(\Alg)\subset\Omega_u^{p+q}(\Alg)$;\\
   iii) $d:\Omega_u^k(\Alg)\rightarrow\Omega_u^{k+1}(\Alg), k=0,1,...$
   is a linear homomorphism satisfying graded Leibnitz rule
   \[
    d(\omega_p\omega^\pr)=d(\omega_p)\omega^\pr+(-)^p\omega_pd(\omega^\pr),
    \forall \omega_p\in\Omega_u^p(\Alg),\omega^\pr\in\Omega_u(\Alg)
   \]
   and nilpotent rule $d\cdot d=0$;\\
   iv) If $\unit$ is the unit of $\Alg$, then $\unit$ is the unit of $\Omega_u(\Alg)$.
   \end{definition}
   Accordingly, one can check that
   \begin{lemma}
    i) $e^xde^y$ for all $x\neq y$ form a linear basis of
    $\Omega_u^1(\Alg)$.\\
    ii) $\chi^x=\sum_{y\in\intg^d}e^yde^{y+x}$ form a module basis of $\Omega_u^1(\Alg)$ which is
    translation-invariant;\\
    iii) (Fundamental noncommutative relation of lattice
    differential):
    \eq\label{1}
     \chi^xf=(T_xf)\chi^x
    \en
    iv) Define a formal partial derivative $\partial_xf(y)=f(y+x)-f(y)=((T_x-\unit)f)(y)$, then there is
    \eq\label{dfU}
     df=\sum_x\partial_xf\chi^x
    \en
    for all $f\in \Alg$.
   \end{lemma}
   The universal differential of function defined in Eq.(\ref{dfU}) is highly non-local in the sense
   that lattice is treated as a spacetime model in physics. So we need a {\it reduction} procedure, namely
   introducing a set of equivalent relation on $\Omega_u(\Alg)$ and
   considering the quotient as {\it differential structure} of this lattice. We will
   use $(\Omega(\Alg),d)$ to denote the quotient differential algebra.\\

   Here begins what we hope to deliver in this paper.
   \begin{definition} (Symmetric Nearest First Order Reduction)\\
    \eq\label{reduction_n}
     df\cong\sum_{\mu=1}^d(\partial_\mu f\chi^\mu +\partial_{-\mu}f\chi^{-\mu})
    \en
    in which $\mu$ is the unit vector along the $\mu$th axis of $\intg^d$
   \end{definition}
   To be compatible with constructive axioms in Definition \ref{def1}, esp. nilpotent rule $d^2=0$,
   relations in order two are inferred.
   \begin{lemma}
    For all $\mu,\nu=1,2,...,d$,\\
    i) Exterior product:
    \eq\label{2}
     \{\chi^{\pm\mu}, \chi^{\pm\nu}\}=0
    \en
    ii) Maurer-Cartan Equation:
    \eq\label{3}
     \{\chi^{\pm\mu}, \chi^{\mp\nu}\}=\dlt^{\mu\nu}d\chi^\mu=\dlt^{\mu\nu}d\chi^{-\nu}
    \en
   \end{lemma}
   Additional to equivalent relation Eq.(\ref{reduction_n}) in first order,
   a set of second order reduction $d\chi^\mu\cong 0\cong d\chi^{-\nu}$ is put into Eq.(\ref{3}), for all
   $\mu,\nu=1,2,...,d$, while the consistency is obvious.
   Consequently, we reach a 2d-dimensional exterior algebra generated by 2d
   translation-invariant 1-forms $\chi^{\pm\mu}$, together with
   noncommutative relation Eq.(\ref{1}) as a {\it canonical differential structure} on $\intg^d$.\\

   The differential structure $(\Omega(\Alg),d)$ which we introduce onto a lattice is able to be represented as a {\it quantized
   calculus} \cite{NCG}.
   \begin{lemma}
    Let $\Hs=\Alg\otimes \complex^{2^d}$ be a complex $l^2$-space defined in the standard
    way; $\Alg$ acts on $\Hs$ by multiplication and the action is
    written as $\pi$. Let
    \[
     \Drc=\sum_\mu(\Gamma^\mu T_\mu +\Gamma^{-\mu} T_{-\mu})
    \]
    in which $\Gamma^{\pm\mu}$ are gamma-matrices in 2d-Euclidean
    space satisfying generating relations of Clifford algebra
    $Cl(E^{2d})$:
    \eq\label{Gamma}
     \{\Gamma^{\pm\mu}, \Gamma^{\pm\nu}\}=0,
     \{\Gamma^{\pm\mu}, \Gamma^{\mp\nu}\}=\dlt^{\mu\nu},
     (\Gamma^\mu)^\dag=\Gamma^{-\mu}
    \en
    and being represented on $\complex^{2d}$ irreducibly.
    Then $(\Hs, \Drc)$ forms a Fredholm Module over $\Alg$.
   \end{lemma}
   In fact, one can verify {\it geometric square-root condition} $\Drc^2=d\unit$,
   hence $\Drc$ is a {\it Fredholm operator} up to a scalar normalization.
   The first step to implement differential representation of $(\Omega(\Alg),d)$ is the introduction of a {\it quantized
   differential}
   \eq\label{quan}
    \hat{d}f=[\Drc, \pi(f)]
   \en
   and the extension of $\pi$ to be a linear homomorphism from $\Omega(\Alg)$ into $End_\complex(\Hs)$ by
   \[
    \pi(f_0df_1df_2...df_p)=\pi(f_0)\cdot\hat{d}f_1\cdot\hat{d}f_2\cdot...\cdot\hat{d}f_p
   \]
   Note that we omit an ``$i$'' in RHS of Eq.(\ref{quan}) which appears in usual literature due
   to the reason that we do not concern the involutive property of differential algebra in
   this work.
   Then, one can check that
   \begin{proposition}(Representation of First Order Reductions)
    \[\pi(\chi^{\pm\mu})=\Gamma^{\pm\mu}T_{\pm\mu}\mbox{ (no
    summation to $\mu$)}
    \]
    \eq\label{map}
     \pi(d\chi^\mu)=\pi(d\chi^{-\nu})=\unit, \forall \mu,\nu=1,2,...,d
    \en
   \end{proposition}
   Eq.(\ref{map}) depicts the common feature for constructing calculus in sNCG
   that differential forms in different orders are mixed. However, this drawback can be cured in our specified model by implement the second order
   reductions, namely define the product of two adjunct gamma matrices to be a wedge
   product. Note importantly that this definition is consistent with Eq.(\ref{1}),
   because of the abelian nature of $\intg^d$.
   Inner product of two forms in $\Omega(\Alg)$ is pulled back
   from the trace of operators on $\Hs$
   \[
    (\omega,\omega^\pr)=Tr(\pi(\omega)^\dag\pi(\omega^\pr)),
    \forall\omega,\omega^\pr\in\Omega(\Alg)
   \]
   in the conventional way of sNCG. One can check that
   the perpendicularity between forms in different order is an
   outcome instead of a prerequisite, thanks to our wedge product definition.
   {\bf Remarks}:\\
    1) It is important to realize that Eqs.(\ref{2})(\ref{3}) are not assumptions, but
    inferences.\\
    2) Second order reductions, though at the first sight appearing to
    be {\it ad hoc} and not so intuitive as the first order ones, are
    necessary for the requirement $\Omega^p(\Alg)\subset
    \Omega^q(\Alg)^\bot, p\neq q$ when inner product of forms is defined.\\
    3) If there be no geometric square root condition, the choice of $\Drc$ is not unique, due to that only $[\Drc, \pi(f)]$ is concerned to
    implement $\Omega(\Alg)$ onto $\Hs$. In fact, $\Drc^\pr=\Drc+{\cal O}$ will do the same work if
    ${\cal O}\in \pi(\Alg)^\pr$ where $\pi(\Alg)^\pr$ is the
    commutants of $\pi(\Alg)$ on $\Hs$. Nevertheless, one has to consider ``{\it
    Junk-idea}'' by using $\Drc^\pr$ to realize differential forms in $End_\complex(\Hs)$, if $\Drc^\pr$ is not Fredholm operator.\\
    4) The distinction between Fredholm operator and Dirac
    operator which has essential implication in operator algebraic
    approach to NCG is not relevant to our stage. In fact, we can
    make a compactification $\intg\rightarrow \intg_N$ with a
    large enough $N$, then we would just handle a finite dimensional
    NCG.
  \section{Staggered Fermions}\label{b}
   Now we make the transition
   \[
    \Drc=\sum_\mu(\Gamma^\mu T_\mu +\Gamma^{-\mu}
    T_{-\mu})\longrightarrow
    \Drc_{dyn}=\sum_\mu(\Gamma^\mu\partial_\mu +\Gamma^{-\mu}\partial_{-\mu})
   \]
   in which $\Drc_{dyn}$ satisfies that
   \begin{proposition}(Physical Square-Root Condition):
   \eq\label{SRP}
    \Drc^2=\Delta
   \en
   where $\Delta=\sum_\mu\partial_\mu\partial_{-\mu}$ is lattice
   Laplacian.
   \end{proposition}
   We will try to show the nontrivial correspondence between $\Drc_{dyn}$ and
   staggered Dirac operator under a specific matrix \repv for $\Gamma^{\pm\mu}$ when $d=1,2,4$
   in this section.\\

   Massless staggered \dopv on d-dimensional lattice, acting on $\Hs_S=\Alg$ directly, can be written as
   \eq\label{Stagg_KS}
    \Drc_S=\sum_\mu\eta^\mu\nabla_\mu
   \en
   in which $\nabla_\mu={1\over 2}(T_\mu-T_{-\mu})$ and
   $\eta^\mu(x)=i(-)^{\sum_{l
   <\mu}x^l}$ is referred as staggered phase \cite{k-s}. Note that our definition of staggered phase has an additional
   ``i'' to insure $\Drc_S$ to be hermitian instead of to be anti-hermitian. Chirality operator is defined to be
   $\epsilon(x)=(-)^{\sum_ix^i}$ \cite{gupta}. The main advantage
   of staggered formalism in LFT is the remnant $U(1)$-chiral
   symmetry generated by $\epsilon(x)$, compared with Wilson-Dirac formalism.
   However, flavor interpretation is a problem for staggered fermion \cite{Stagg_Flv}. When a ``double spacing'' transformation being performed,
   staggered Dirac operator in Susskind form as in Eq.(\ref{Stagg_KS}) could converted into a bi-module form to which the right module is
   interpreted as flavor space; this equivalent is broken when a
   gauge potential presents on lattice. We will adopt this
   ``double spacing'' tech also below. Dynamics of
   staggered fermions and gauge fields has been well studied in
   \cite{KS}.\\

   In the present work, staggered fermion field is denoted as
   $\phi$ whose classical action functional is $A[\phi]=(\phi, \Drc_S\phi)_{\Hs_S}=\sum_x\phi(x)^\ast(\Drc_S\phi)(x)$.
   As for our formalism, fermion
   fields are elements in $\Hs$, being written as $\psi$ with $2^d$-components; classical action is
   $A[\psi]=(\psi, \Drc_{dyn}\psi)_\Hs=\sum_x\psi^\dag(x)(\Drc_{dyn}\psi)(x)$. The subtlety concerning anti-commutativity for Euclidean spinor
   is not relevant in this work, so we do not use the notation like $\overline{\phi},\overline{\psi}$.
   \begin{proposition}
    $A[\phi]=A[\psi]$ when $d=1,2,4$.
   \end{proposition}
   \prf
   {\bf d=1:} We modify the \repv used by Dimakis and M\"{u}ller-Hoissen in
   \cite{dimakis1} to be that
   $\Gamma^1=\left( \begin{array}{cc}0&0\\i&0\end{array}\right)$,
   $\Gamma^{-1}=(\Gamma^1)^\dag$, and introduce a ``double-spacing''
   lattice by defining the map $\psi_1(x)=\phi(2x)/\sqrt{2},
   \psi_2(x)=\phi(2x+1)/\sqrt{2}$, then one can check $A[\psi]=A[\phi]$.\\

   {\bf d=2:} Let
   \[
    \Gamma^{(1,0)}=\left(\begin{array}{cccc}
    0&0&0&-i\\0&0&-i&0\\0&0&0&0\\0&0&0&0
    \end{array}\right),
    \Gamma^{(0,1)}=\left(\begin{array}{cccc}
    0&-i&0&0\\0&0&0&0\\0&0&0&0\\0&0&i&0
    \end{array}\right)
   \]
   $\Gamma^{(-1,0)}=(\Gamma^{(1,0)})^\dag,
   \Gamma^{(0,-1)}=(\Gamma^{(0,1)})^\dag$, and let ``double-spacing''
   map be $\psi_3(x^1,x^2)=\phi(2x^1,2x^2)/\sqrt{2}$, $\psi_2(x^1,x^2)=\phi(2x^1+1,2x^2)/\sqrt{2}$,
   $\psi_4(x^1,x^2)=\phi(2x^1,2x^2+1)/\sqrt{2}$, $\psi_1(x^1,x^2)=\phi(2x^1+1,2x^2+1)/\sqrt{2}$. Then
   $A[\psi]=A[\phi]$ still holds. Note that $\epsilon$ being mapped onto $\Hs$
   equals to $diag(1,-1,1,-1)$.\\

   {\bf d=4:} Label spinor components of $\psi$ by
   $\psi_{\hat\dlt}$ in which
   $\hat{\dlt}=(\dlt^1,\dlt^2,\dlt^3,\dlt^4),
   \dlt^i\in\{0,1\},i=1,...,4$, and order $\hat{\dlt}$ as
   $(0,0,0,0)$, $(0,0,1,1)$, $(0,1,0,1)$, $(1,0,0,1)$, $(0,1,1,0)$, $(1,0,1,0)$, $(1,1,0,0)$, $(1,1,1,1)$,
   $(0,1,1,1)$, $(1,0,1,1)$, $(1,1,0,1)$, $(1,1,1,0)$, $(0,0,0,1)$, $(0,0,1,0)$, $(0,1,0,0)$,
   $(1,0,0,0)$. Under this ordering, define the \repv of $\Gamma^{(1,0,0,0)}$, $\Gamma^{(0,1,0,0)}$,
   $\Gamma^{(0,0,1,0)}$, $\Gamma^{(0,0,0,1)}$ to be
   \footnotesize
   \[
    \left(
    \begin{array}{cccccccccccccccc}
     0&0&0&0&0&0&0&0&0&0&0&0&0&0&0&0\\
     0&0&0&0&0&0&0&0&0&0&0&0&0&0&0&0\\
     0&0&0&0&0&0&0&0&0&0&0&0&0&0&0&0\\
     0&0&0&0&0&0&0&0&0&0&0&0&i&0&0&0\\
     0&0&0&0&0&0&0&0&0&0&0&0&0&0&0&0\\
     0&0&0&0&0&0&0&0&0&0&0&0&0&i&0&0\\
     0&0&0&0&0&0&0&0&0&0&0&0&0&0&i&0\\
     0&0&0&0&0&0&0&0&i&0&0&0&0&0&0&0\\
     0&0&0&0&0&0&0&0&0&0&0&0&0&0&0&0\\
     0&i&0&0&0&0&0&0&0&0&0&0&0&0&0&0\\
     0&0&i&0&0&0&0&0&0&0&0&0&0&0&0&0\\
     0&0&0&0&i&0&0&0&0&0&0&0&0&0&0&0\\
     0&0&0&0&0&0&0&0&0&0&0&0&0&0&0&0\\
     0&0&0&0&0&0&0&0&0&0&0&0&0&0&0&0\\
     0&0&0&0&0&0&0&0&0&0&0&0&0&0&0&0\\
     i&0&0&0&0&0&0&0&0&0&0&0&0&0&0&0
    \end{array}\right),
   \]
   \[
    \left(
    \begin{array}{cccccccccccccccc}
     0&0&0&0&0&0&0&0&0&0&0&0&0&0&0&0\\
     0&0&0&0&0&0&0&0&0&0&0&0&0&0&0&0\\
     0&0&0&0&0&0&0&0&0&0&0&0&i&0&0&0\\
     0&0&0&0&0&0&0&0&0&0&0&0&0&0&0&0\\
     0&0&0&0&0&0&0&0&0&0&0&0&0&i&0&0\\
     0&0&0&0&0&0&0&0&0&0&0&0&0&0&0&0\\
     0&0&0&0&0&0&0&0&0&0&0&0&0&0&0&-i\\
     0&0&0&0&0&0&0&0&0&-i&0&0&0&0&0&0\\
     0&i&0&0&0&0&0&0&0&0&0&0&0&0&0&0\\
     0&0&0&0&0&0&0&0&0&0&0&0&0&0&0&0\\
     0&0&0&-i&0&0&0&0&0&0&0&0&0&0&0&0\\
     0&0&0&0&0&-i&0&0&0&0&0&0&0&0&0&0\\
     0&0&0&0&0&0&0&0&0&0&0&0&0&0&0&0\\
     0&0&0&0&0&0&0&0&0&0&0&0&0&0&0&0\\
     i&0&0&0&0&0&0&0&0&0&0&0&0&0&0&0\\
     0&0&0&0&0&0&0&0&0&0&0&0&0&0&0&0
    \end{array}\right),
   \]
   \[
    \left(
    \begin{array}{cccccccccccccccc}
     0&0&0&0&0&0&0&0&0&0&0&0&0&0&0&0\\
     0&0&0&0&0&0&0&0&0&0&0&0&i&0&0&0\\
     0&0&0&0&0&0&0&0&0&0&0&0&0&0&0&0\\
     0&0&0&0&0&0&0&0&0&0&0&0&0&0&0&0\\
     0&0&0&0&0&0&0&0&0&0&0&0&0&0&-i&0\\
     0&0&0&0&0&0&0&0&0&0&0&0&0&0&0&-i\\
     0&0&0&0&0&0&0&0&0&0&0&0&0&0&0&0\\
     0&0&0&0&0&0&0&0&0&0&i&0&0&0&0&0\\
     0&0&-i&0&0&0&0&0&0&0&0&0&0&0&0&0\\
     0&0&0&-i&0&0&0&0&0&0&0&0&0&0&0&0\\
     0&0&0&0&0&0&0&0&0&0&0&0&0&0&0&0\\
     0&0&0&0&0&0&i&0&0&0&0&0&0&0&0&0\\
     0&0&0&0&0&0&0&0&0&0&0&0&0&0&0&0\\
     i&0&0&0&0&0&0&0&0&0&0&0&0&0&0&0\\
     0&0&0&0&0&0&0&0&0&0&0&0&0&0&0&0\\
     0&0&0&0&0&0&0&0&0&0&0&0&0&0&0&0
    \end{array}\right),
   \]
   \[
    \left(
    \begin{array}{cccccccccccccccc}
     0&0&0&0&0&0&0&0&0&0&0&0&0&0&0&0\\
     0&0&0&0&0&0&0&0&0&0&0&0&0&-i&0&0\\
     0&0&0&0&0&0&0&0&0&0&0&0&0&0&-i&0\\
     0&0&0&0&0&0&0&0&0&0&0&0&0&0&0&-i\\
     0&0&0&0&0&0&0&0&0&0&0&0&0&0&0&0\\
     0&0&0&0&0&0&0&0&0&0&0&0&0&0&0&0\\
     0&0&0&0&0&0&0&0&0&0&0&0&0&0&0&0\\
     0&0&0&0&0&0&0&0&0&0&0&-i&0&0&0&0\\
     0&0&0&0&i&0&0&0&0&0&0&0&0&0&0&0\\
     0&0&0&0&0&i&0&0&0&0&0&0&0&0&0&0\\
     0&0&0&0&0&0&i&0&0&0&0&0&0&0&0&0\\
     0&0&0&0&0&0&0&0&0&0&0&0&0&0&0&0\\
     i&0&0&0&0&0&0&0&0&0&0&0&0&0&0&0\\
     0&0&0&0&0&0&0&0&0&0&0&0&0&0&0&0\\
     0&0&0&0&0&0&0&0&0&0&0&0&0&0&0&0\\
     0&0&0&0&0&0&0&0&0&0&0&0&0&0&0&0
    \end{array}\right)
   \]
   \normalsize
   respectively, and $\Gamma^{-\mu}=(\Gamma^\mu)^\dag$;
   let ``double-spacing'' map to be $\psi_{\hat{\dlt}}(x)=\phi(2x+\hat{\dlt})/\sqrt{2}$.
   After a tedious algebra, one will reach still
   $A[\psi]=A[\phi]$; while $\epsilon=diag(\unit_8,-\unit_8)$.\\
   \endprf
   We would like to conjecture that
   {\it there exists a \repv for $\Gamma^\mu, \Gamma^{-\mu}, \mu=1,2,...,d$ and a ``double-spacing'' map from $\Hs_S$ to
   $\Hs$, such that $A[\psi]=A[\phi]$ holds for any $d$.}\\

   {\bf Remarks}:\\
   1) Square root property Eq.(\ref{SRP}) which is emphasized by Vaz in \cite{v} has been ignored by previous authors \cite{Stagg_Flv}.
   In our understanding, this property characterizes staggered Dirac operator in an abstract sense.\\
   2) Our \DCv operator can be understood also as an abstract
   definition
   of staggered operator where ``abstract'' refers to representation independent.
  \section{Discussions}\label{c}
   Some similarity in formalism can be found in the work of Takami \etal
   \cite{japan}. In fact, they were considering a discretized
   Weyl-equation on lattice in their papers. Combine a discrete
   time axis ${\cal T}$ to the above $\intg^d$, and define forward action of
   translation along ${\cal T}$ to be $(T_0^+f)(t,x)=f(t+1,x)$ and
   $\partial_tf=T_0^+f-f$. Then their \dopv $\Lambda$ can be essentially
   written as
   \[
    \Lambda=-\partial_t+T_0^+D
   \]
   where $D$ is just $\Drc_{dyn}$, \DCv operator discussed in the last section. $\Lambda$ is not hermitian, though
   these author showed that this lost would not do harm to physics.\\


   {\bf Acknowledgements}\\
    This work was supported by Climb-Up (Pan Deng) Project of
    Department of Science and Technology in China, Chinese
    National Science Foundation and Doctoral Programme Foundation
    of Institution of Higher Education in China. We are grateful
    for Prof. A. Rivero for the help on mathematics of Fredholm module.

  
 \end{document}